\documentclass[]{spie}  %>>> use for US letter paper
\usepackage{amsmath,amsfonts,amssymb}
\usepackage{graphicx}
\usepackage[colorlinks=true, allcolors=blue]{hyperref}
\usepackage{lipsum}
\usepackage{gensymb}
\usepackage{float}

\title{HARMONI at ELT: A Zernike wavefront sensor for the high-contrast module - Testbed results with realistic observation conditions}

\author[a]{Adrien Hours}
\author[a]{Alexis Carlotti}
\author[a]{David Mouillet}
\author[a]{Alain Delboulbé}
\author[a]{Sylvain Guieu}
\author[a]{Laurent Jocou}
\author[a]{Thibaut Moulin}
\author[a]{Fabrice Pancher}
\author[a]{Patrick Rabou}
\author[b]{Elodie Choquet}
\author[b]{Kjetil Dohlen}
\author[b,c]{Jean-François Sauvage}
\author[d]{Mamadou N'Diaye}
\affil[a]{Univ. Grenoble Alpes, CNRS, IPAG, 38000 Grenoble, France}
\affil[b]{Aix Marseille Université, CNRS, LAM (Laboratoire d\textquotesingle Astrophysique de Marseille) UMR 7326, 13388 Marseille, France}
\affil[c]{ONERA, The French Aerospace Lab, 92322 Châtillon, France}
\affil[d]{Université Côte d\textquotesingle Azur, Observatoire de la Côte d'Azur, CNRS, Laboratoire Lagrange, 06108 Nice, France}

\authorinfo{Further author information: (Send correspondence to Adrien Hours)\\Adrien Hours: E-mail: adrien.hours@univ-grenoble-alpes.fr\\}

\begin{document} 
\maketitle

\begin{abstract}
ELT-HARMONI is the first light visible and near-IR integral field spectrograph (IFS) for the ELT. It covers a large spectral range from 450nm to 2450nm with resolving powers from 3500 to 18000 and spatial sampling from 60mas to 4mas. It can operate in two Adaptive Optics modes - SCAO (including a High Contrast capability) and LTAO - or with NOAO. The project is preparing for Final Design Reviews.

\noindent
The High Contrast Module (HCM) will allow HARMONI to perform direct imaging and spectral analysis of exoplanets up to \(10^6\) times fainter than their host star. Quasi-static aberrations are a limiting factor and must be calibrated as close as possible to the focal plane masks to reach the specified contrast. A Zernike sensor for Extremely Low-level Differential Aberrations (ZELDA) will be used in real-time and closed-loop operation at 0.1Hz frequency for this purpose. Unlike a Shack-Hartmann, the ZELDA wavefront sensor is sensitive to Island and low-wind effects. The ZELDA sensor has already been tested on VLT-SPHERE\cite{Mamadou2016} and will be used in other instruments. Our objective is to adapt this sensor to the specific case of HARMONI.

\noindent
A ZELDA prototype is being both simulated and experimentally tested at IPAG. Its nanometric precision has first been checked in 2020 in the case of slowly evolving, small wavefront errors, and without dispersion nor turbulence residuals. On this experimental basis, we address the performance of the sensor under realistic operational conditions including residuals, mis-centring, dispersion, sensitivity, etc. Atmospheric refraction residuals were introduced by the use of a prism, and turbulence was introduced by a spatial light modulator which is also used to minimise wavefront residuals in a closed loop in the observing conditions expected with HARMONI.
\end{abstract}


\section{INTRODUCTION}
\label{sec:intro}  % \label{} allows reference to this section

%Pour quelles raisons on choisit d'avoir un zelda dans harmoni ? (SCAO a un lambda, ifs a un autre, optiques differentielles entre les deux)
%HARMONI : presenter ses objectifs et ses contraintes auxquelles on doit faire face (et dans ce cadre, avoir un senseur dedie, ici le zelda)
%Pk on travaille la dessus? qu'est-ce qu'on apporte de plus par rapport a ce qui a deja ete fait? 1- on a des residus de dispersion et le ZELDA n'est pas concu pour travailler avec de tels residus 2- les residus AO de harmoni ne seront pas les memes que ceux de sphere
%on aborde des challenges, c'est pas juste une reapplication de la meme methodologie ou on a juste change les chiffres, bien souligner qu'il y a des defis et que ca demande une reflexion specifique et qu'il y a une validation test (faire ca le plus tot possible, dans l'abstract ?). Aspect theorique, expliquer comment ca marche : etre assez bref 

Many exoplanets have already been discovered using indirect detection techniques such as transits, radial velocities, astrometry or gravitational microlensing. However, in order to make a complete spectral analysis of the atmosphere of these exoplanets, they must be observed by direct imaging. 

\noindent
Exoplanet direct imaging is challenging: the exoplanet's faint signature at very short separation is hidden within the stellar diffracted light. It is however possible to improve the signal to noise ratio by i/ measuring and correcting the aberrations due to atmospheric turbulence, and unseen optical defects (Non Common Path Aberrations, hereafter NCPA) with adaptive optics, and by ii/ decreasing the intensity of the diffracted light of the star with a coronagraph which removes this diffracted light out of the star's neighborhood with pupil apodization. Finally, post processing techniques aim at identifying and discriminating the remaining aberration signatures from the signal of interest\cite{carlotti:hal-02118132}.

\noindent
HARMONI (High Angular Resolution Monolithic Optical and Near-infrared Integral field spectrograph) will be ESO's ELT first-light visible and near infrared spectro-imager. As a general-purpose instrument, HARMONI will cover a wide range of science cases, with various observing modes. This study is motivated by the High contrast Module (HCM) which focusses on the NIR red part for which HARMONI have a maximum \(R=18000\) resolving power. The HCM is designed to give HARMONI the capability to observe exoplanets that are located as close as 0.1'' and that have a flux ratio down to $10^{-6}$ relative to the host star.

\noindent
The Single Conjugate Adaptive Optics (SCAO) subsystem will be used to provide a 75-80\% Strehl ratio for a median seeing (0.65'') in the K-band. The correction of the wavefront will be ensured by the M4 deformable mirror of the telescope after being measured by a pyramid wavefront sensor at 0.8 $\mu$m. The IFS will operate from 1.2 to 2.4 $\mu$m. The difference between the sensing and the observing wavelengths will induce NCPA. To attenuate them, a ZELDA wavefront sensor\cite{Mamadou2013} will be part of the high-contrast module (HCM) of HARMONI, which is located between the SCAO subsystem, and the IFS. 

%There is thus a difference in wavelength to which are added residues of adaptive optics (because the SCAO is not an extreme adaptive optics system as it is the case for SPHERE) and residues of atmospheric dispersion.  ACa : Non, ça n'est pas le fait que la SCAO de HARMONI est moins bonne, c'est surtout que les faisceaux passent à différents endroits des optiques en fct des longueurs d'onde, et donc qu'il est vu différemment par le senseur et par l'IFS. Dans SPHERE ça avait poussé David à dire à Patrick de donner une super qualité aux optiques loin du plan pupille. On fait pareil dans HARMONI.

\noindent
The \ref{sec:part_2}\(^{nd}\) section recalls the basic principle of the ZELDA, and details some specific challenges to be addressed in the operational case of HARMONI. Section \ref{sec:part_3} presents the optical bench used to test the ZELDA sensor under the expected typical HARMONI observing conditions by reproducing them experimentally, and the results obtained so far. The \ref{sec:conclu}\(^{th}\) and final section draws some perspectives and a conclusion.

%%% PART 2 %%%
\section{THE ZELDA WAVEFRONT SENSOR IN THE SPECIFIC CASE OF HARMONI}
\label{sec:part_2}

The objective of the ZELDA wavefront sensor is to measure the quasi-static aberrations introduced by the differential optics between the SCAO module and the IFS. These Non-Common Path Aberrations (NCPA) must be measured and corrected with an accuracy of less than 5 nm rms for stars with magnitudes ranging from -4 to 12. Measuring frequency must be between 0.06 and 0.1 Hz and within a range of $\pm 60$ nm rms.

\noindent
The ZELDA wavefront sensor has already been used with VLT-SPHERE\cite{Mamadou2016} in which it is an engineering mode. In HARMONI, however, it will be used to assist all high-contrast observations.
%visiting instrument but needs to be tested and validated under the specific conditions of HARMONI as a permanent instrument. 
The first difference between VLT-SPHERE and HARMONI is the shape of the telescope aperture, the influence of which is detailed in subsection \ref{sec:part_2_2}. 

\noindent
The second major difference is that unlike in VLT-SPHERE, the SCAO subsystem of HARMONI will not be an extreme AO system, an residual atmospheric turbulence will interfere more strongly with the wavefront measurements. %là encore l'idée ça n'est pas qu'avec SPHERE il n'y a plus de résidus d'AO, c'est qu'ils sont plus faibles qu'avec HARMONI. SPHERE n'est pas parfait. Aucun système ne l'est ou ne le sera. L'important c'est de connaitre le niveau d'erreur, et de construire le système pour celui-ci.

\noindent
Finally, HARMONI does not correct by default for the atmospheric dispersion. A passive atmospheric dispersion corrector (ADC) is however part of the HCM, and it is optimized for observations at a zenith distance of 32\degree. The HCM will observe in a 5-50$\degree \ $zenith distance range, and there will therefore be atmospheric dispersion residuals that will impact the wavefront measurements. %Je corrige des trucs dont on a discuté lors de la préparation de ta présentation : le range de 5-50 est un choix pour le HCM. L'ELT peut aller au-delà. Il n'y a aucun ADC pour tout HARMONI, mais on a choisi d'en mettre un dans le HCM.

%HARMONI conditions:
%\begin{itemize}
%    \item ZELDA specs (range, accuracy, ...)
%    \item Goal: quasi-static aberrations NCPA measurements.
%    \item SCAO \(\Rightarrow\) AO residuals.
%    \item ADC \(\Rightarrow\) Atmosphere refraction residuals max at 5 or 50 degree zenith distance.
%\end{itemize}

\subsection{Physics Principle}
\label{sec:part_2_1}

The ZELDA wavefront sensor\cite{Mamadou2013} consists of a phase mask with a cylindrical shape etched into a glass window and placed in the focal plane of a lens. The diameter of the mask is equal to the full width at half maximum of an Airy disk and its depth is such that the introduced phase shift $\theta$ is close to $\frac{\pi}{2}$. As a result it converts the phase variations in the entrance pupil plane $\varphi_{\text{in}}$ into intensity variations in the exit pupil plane $I_{\text{out}}$. If the correction of the aberrations due to atmospheric turbulence is good enough, i.e., if the Strehl ratio is high enough, the following relation can be derived:

\begin{equation}
I_{\text{out}} = P^2 +2b^2(1-\cos\theta) + 2Pb\left[ \sin\varphi_{\text{in}} \sin\theta - \cos\varphi_{\text{in}}(1-\cos\theta) \right]
\label{eqIphi}
\end{equation}

\noindent
where $P$ is the real amplitude of the electric field in the the entrance pupil plane and $b$ the diffraction term defined by:

\begin{equation}
b = \sqrt{S}\widehat{M}\otimes P_0
\label{terme_b}
\end{equation}

\noindent
where $S = \exp{- \sigma_{\varphi}^2} \approx (1 - \sigma_{\varphi}^2/2)^2$ is the Strehl ratio in a low-phase regime with $\sigma_{\varphi}^2$ the wavefront variance, $\widehat{M}$ the Fourier transform of $M$ the top-hat function of the phase mask (equal to 1 for $\vert \rho \vert < d/2$ and 0 otherwise, $d$ the diameter of the mask), and $P_0$ the telescope aperture shape (equal to 1 inside the pupil and 0 elsewhere).

\noindent
As the phase errors are very small, a second order approximation can be made:

\begin{align}
\begin{split}
    I_{\text{out}} &= P^2 +2b^2(1-\cos\theta) + 2Pb\left[ \varphi_{\text{in}} \sin\theta - (1 - \varphi_{\text{in}}^2/2)(1-\cos\theta) \right] \\
    &= P^2 +2b^2 + 2Pb (\varphi_{\text{in}}^2/2 + \varphi_{\text{in}} - 1)\ \ \ \text{as } \theta=\pi/2 
\label{eqIphi2}
\end{split}
\end{align}

\noindent
Thanks to this approximation, the equation can be inverted to derive the phase from the intensity:

\begin{equation}
\varphi_{\text{in}} = P \left( 1 - \sqrt{3 - \frac{2b}{P} - \frac{1}{b} \left(\frac{P-I_{\text{out}}}{P}\right)} \right) 
\label{eqphi}
\end{equation}

\begin{figure}[H]
	\begin{center}
    	\includegraphics[width=\textwidth]{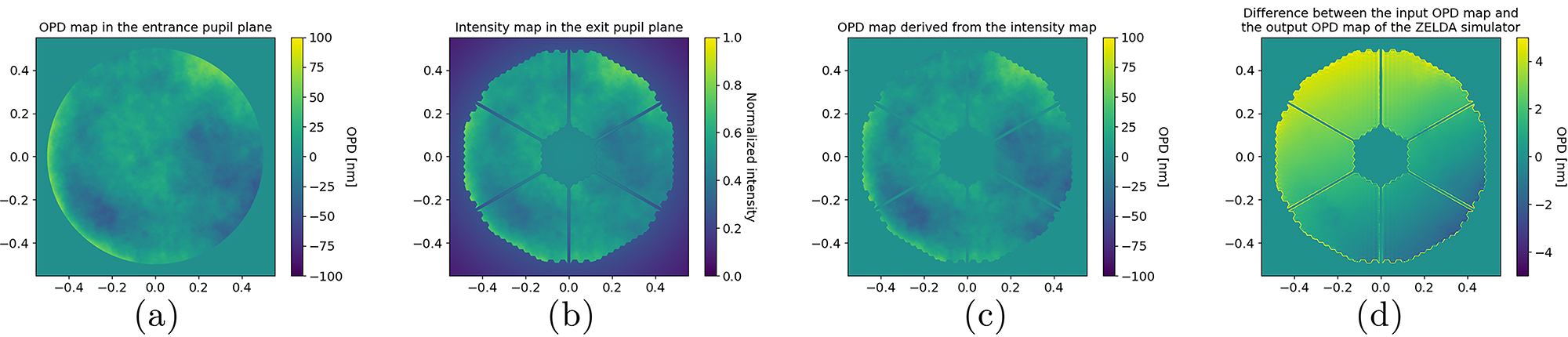}
	\end{center}
	\caption[Description liste figures]{\label{Modele_Zelda} (a) Input OPD map with $f^{-3}$ power law. (b) Corresponding intensity map as seen in the pupil plane after the Zelda. (c) Reconstructed OPD map through equation \ref{eqphi}. (d) Difference between reconstructed and input maps within the pupil.}
\end{figure}

\noindent
The Optical Path Delay (OPD) map can then be derived from the relation $\text{OPD} = \lambda\varphi_{\text{in}} / 2\pi$. An example is given in figure \ref{Modele_Zelda}, with an input map with aberrations adopting a Power Spectral Density (PSD) that follows a spacial frequency $f^{-3}$ power law.  For this case, we have the results summarized in table \ref{Table1}. Wavefront error residuals between 0 and 40 $\lambda/D$ are due to the quadratic approximation but are small enough to be neglected. %ON dit IN figure, pas BY figure. Idem pour les tables, ou les sections.

\begin{table}[H]
\centering
\begin{tabular}{|r|c|c|c|c|} 
\hline
Spatial frequencies region & 0-4 $\lambda$/$D$ & 4-40 $\lambda$/$D$ & 40-100 $\lambda$/$D$ &  ``all'' (0-512 $\lambda$/$D$)  \\ 
\hline
$\varphi_{\text{out}}$ [nm rms]    & 10.91        & 7.30         & 2.22           & 13.88                       \\
\hline
$\Delta\varphi_{\text{out-in}}$ [nm rms]    & 0.35        & 0.36         & 0.46           & 3.80                       \\
\hline
\end{tabular}
\caption[Description liste tableaux]{\label{Table1} Aberrations according to spatial frequencies in the input phase map ($\varphi_{\text{out}}$) and difference between reconstructed and input maps ($\Delta\varphi_{\text{out-in}}$).}
\end{table}

\noindent
As we can see, the model introduces some error in the high spatial frequencies (due to some interpixel misalignment), but as the ELT correcting mirror M$_4$ is an 80 per 80 actuators Deformable Mirror, the wavefront correction will be limited at 40 $\lambda$/$D$.

\subsection{Pupil Shape Effects}
\label{sec:part_2_2}

\begin{figure}[H]
	\begin{center}
    	\includegraphics[width=\textwidth]{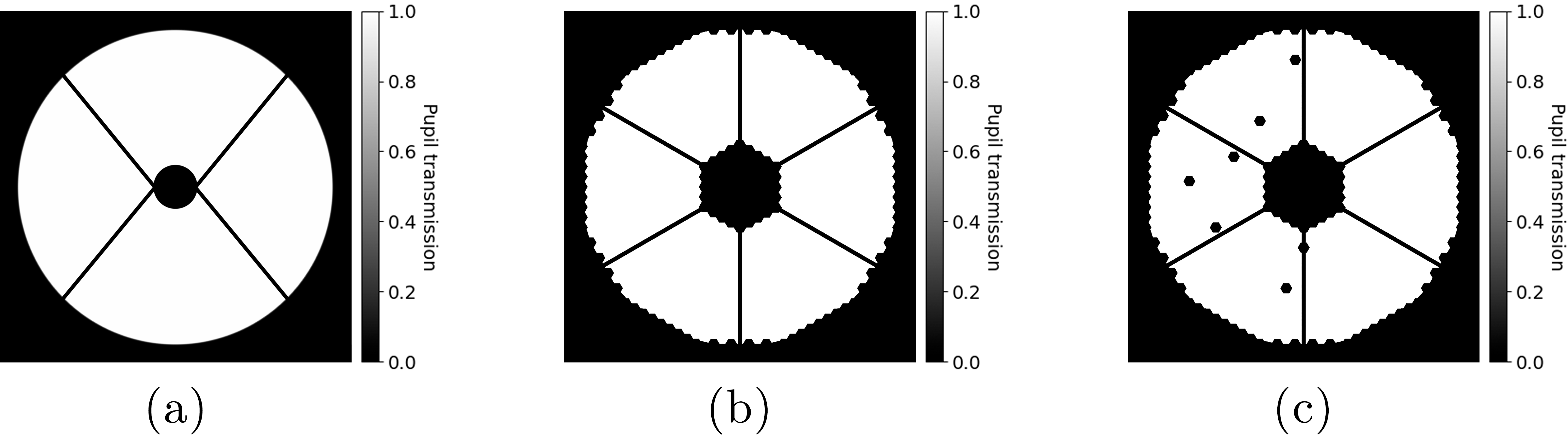}
	\end{center}
	\caption[Description liste figures]{\label{Pupils}The VLT pupil shape (a) and the ELT pupil shape without (b) and with (c) missing segments.}
\end{figure}

As mentioned before, the first difference between the tests performed on VLT-SPHERE and the specific case of HARMONI in which this ZELDA wavefront sensor will work is the shape of the telescope aperture $P_0$, which influences the parameter $b$ (see equation \ref{terme_b}). In the case of VLT-SPHERE, the pupil was circular with a central obstruction supported by four spiders. In the case of HARMONI, the primary mirror will be subdivided into 798 segments distributed in a hexagonal shape with a central obstruction supported by six spiders with potential missing segments (engine failure, mirror re-aluminizing, etc.)  as shown in figure \ref{Pupils}. These missing segments (whose distribution will be random) will introduce interference patterns that will have to be calibrated in post-processing. 

\begin{figure}
	\begin{center}
    	\includegraphics[scale=0.16]{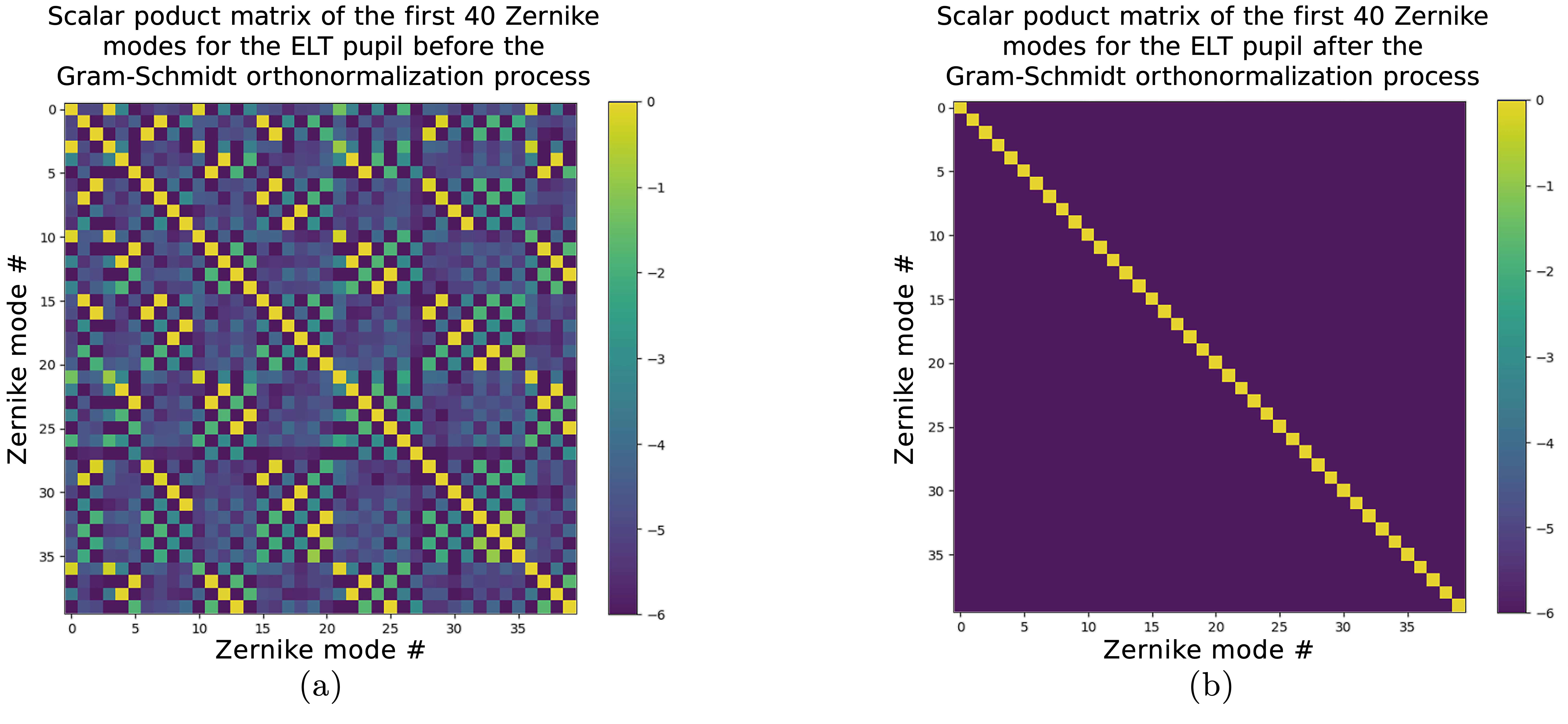}
	\end{center}
	\caption[Description liste figures]{\label{GramSchmidt} Scalar product matrices of the first 40 modes (in \emph{logscale}) (a) before and (b) after using the Gram-Schmidt orthonormalization process.}
\end{figure}

\noindent
This raises two issues:
\begin{itemize}
    \item The basis of the Zernike polynomials is orthogonal for a ciruclar pupil but not for a pupil of arbitrary shape. Therefore, the Gram-Schmidt orthonormalization process must be performed to fit the basis to the shape of the ELT pupil, as shown in figure \ref{GramSchmidt}. %J'ai un peu du mal avec cette justification parce que dès que la pupille n'est pas juste un disque plein, il faut utiliser autre chose que des Zernikes. On le fait déjà sur SPHERE.
    \item The segmentation into 798 mirrors induces a non-negligible probability of having an incomplete pupil either because the motors of some segments will fail prior to the observations, thus leading to segments having to be manually pointed off the optical axis, or because segments will be physically removed from the telescope to be recoated. This results in an ELT pupil with missing segments whose distribution varies from one night to another. These missing segments will create an interference pattern on the Point Spread Function (PSF) of the pupil that may have to be taken into account in post-processing.%, as shown in figure \ref{MissingSegments}. 
\end{itemize}

%\begin{figure}
%	\begin{center}
%    	\includegraphics[scale=0.35]{Images/PSF_ELT_all.png}
%	\end{center}
%	\caption[Description liste figures]{\label{MissingSegments} Missing Segments induce interference pattern.}
%\end{figure}

\begin{figure}[H]
	\begin{center}
    	\includegraphics[width=\textwidth]{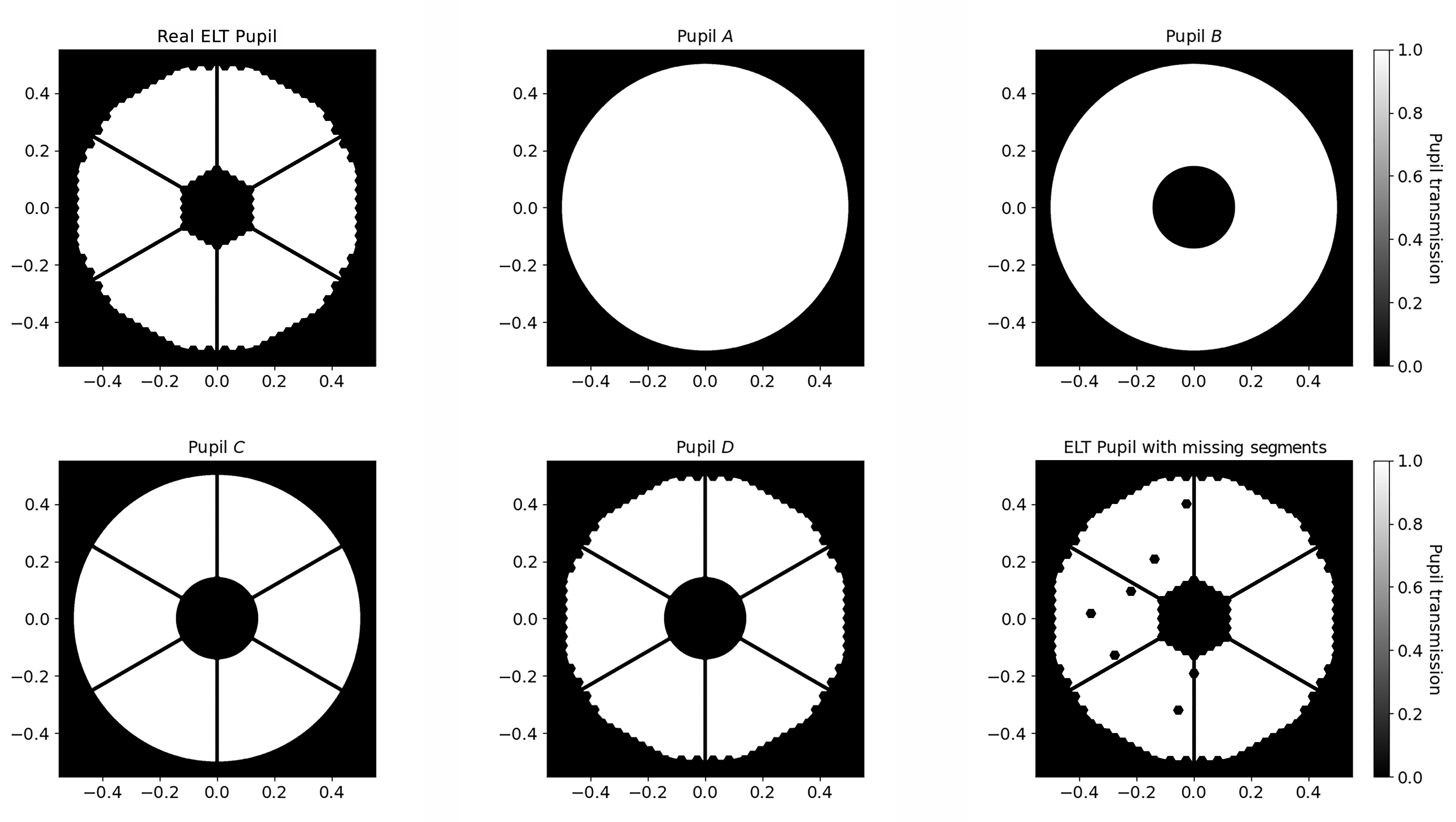}
	\end{center}
	\caption[Description liste figures]{\label{Pupil_approx} The real ELT aperture shape and erroneous aperture shapes considered.}
\end{figure}

\noindent
Moreover, equation \ref{terme_b} shows the dependence of the $b$ term on the telescope aperture shape $P_0$. An error in the pupil shape model can thus result in an error on the $b$ term estimation and eventually on the measured phase estimation. In order to measure the impact of such pupil error, we tested different pupils shown in figure \ref{Pupil_approx}: for each of these pupils, we calculated the associated $b$ term, and then estimated the phase from the same intensity map (figure \ref{Modele_Zelda}(b)) with each of these erroneous $b$ terms. The comparison of each of these phase maps with the phase map (figure \ref{Modele_Zelda}(c)) is given in table \ref{Table2}.

\begin{table}[H]
\centering
\begin{tabular}{|r|c|c|c|c|} 
\hline
Spatial frequencies region & 0-4 $\lambda$/$D$ & 4-40 $\lambda$/$D$ & 40-100 $\lambda$/$D$ & ``all'' (0-512 $\lambda$/$D$)  \\ 
\hline
$\varphi_{\text{out}}$ [nm rms]    & 10.91        & 7.30         & 2.22           & 13.88                       \\
\hline
$\Delta\varphi_{\text{out-A}}$ [nm rms]    & 2.31        & 1.25         & 0.39           & 2.76                       \\
\hline
$\Delta\varphi_{\text{out-B}}$ [nm rms]    & 0.93        & 0.51         & 0.16           & 1.12                       \\
\hline
$\Delta\varphi_{\text{out-C}}$ [nm rms]    & 0.34        & 0.19         & 0.06           & 0.41                       \\
\hline
$\Delta\varphi_{\text{out-D}}$ [nm rms]    & 0.27        & 0.15         & 0.05           & 0.33                       \\
\hline
$ \Delta\varphi_{\text{out-ELT\(_{\text{Missing Segments}}\) }} $ [nm rms]    & 0.16        & 0.09         & 0.03           & 0.19                       \\
\hline
\end{tabular}
\caption[Description liste tableaux]{\label{Table2} Aberrations according to spatial frequencies in the input phase map ($\varphi_{\text{out}}$) and difference between reconstructed and input maps for each of the erroneous $b$ terms.}
\end{table}

\noindent
As can be seen, the dominant impact is the overall shape and central obstruction. Spiders and crenellated edges also play a significant, though weaker, role. The effects of missing segments are negligible.

\noindent
Taking a wrong pupil (by considering a full pupil or neglecting the central obstruction) leads to errors that are too large for the accuracy we want to reach.

\noindent
On the other hand, approximating the central obstruction by a circle (as we will do experimentally) or neglecting the missing segments is acceptable since the errors introduced are subnanometric. 

\noindent
For the rest of this study, we will then use the pupil \textit{D} of figure \ref{Pupil_approx}. Missing segments will also be neglected as they introduces only very small errors in the phase estimation. %Pas besoin de mettre des majuscules à figure ou table, sauf si c'est le premier mot

\subsection{Effects of Atmospheric Dispersion Residuals}
\label{sec:part_2_3}

\begin{figure}[H]
	\begin{center}
    	\includegraphics[scale=0.3]{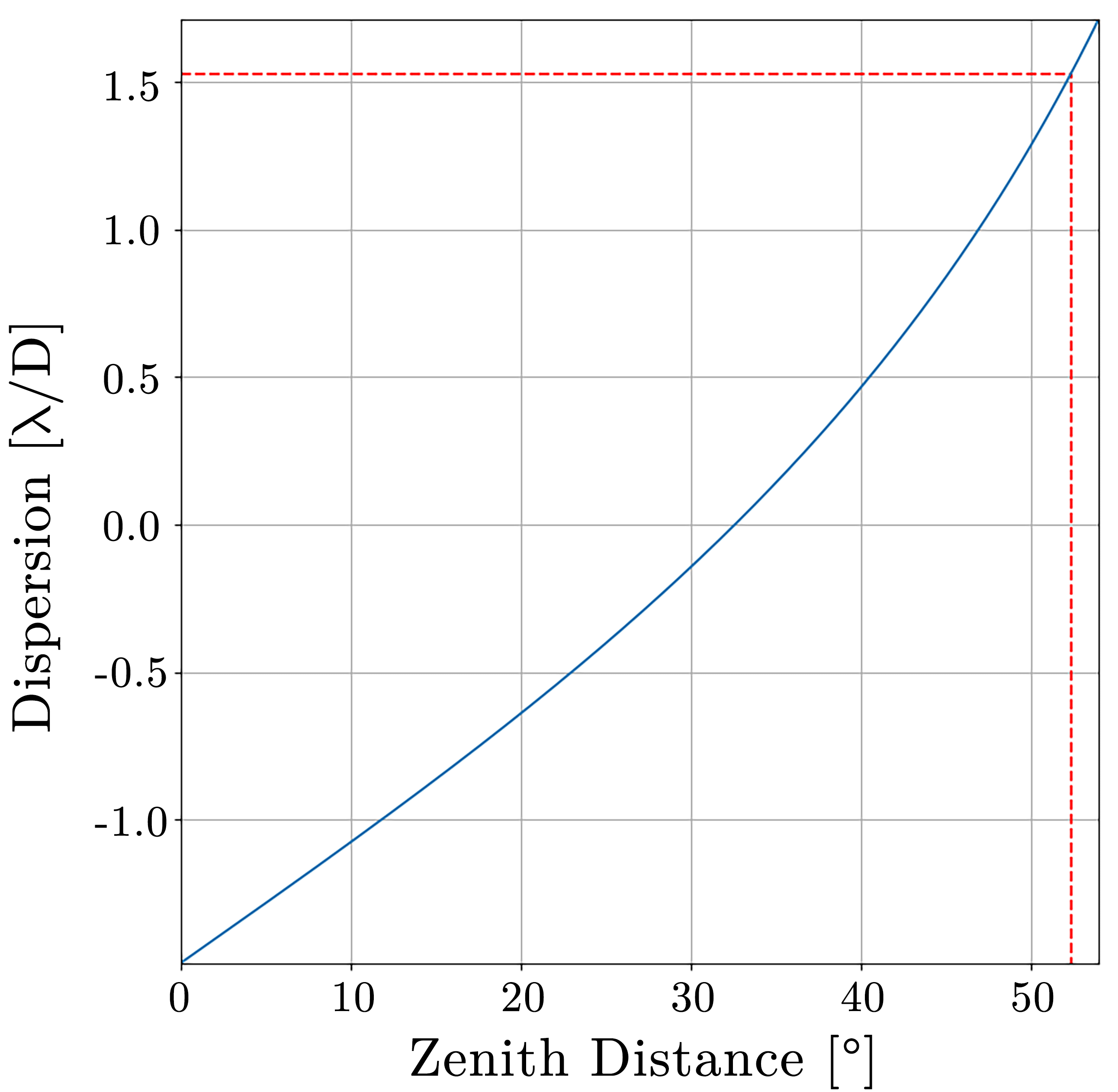}
	\end{center}
	\caption[Description liste figures]{\label{Dispersion} Evolution of the dispersion according to the zenith distance of observation. In order to test the ZELDA in the conservative case beyond the limits of the HCM, and for material reasons that will be detailled in the subsection \ref{sec:part_3_1}, we work in the case of a zenith distance of 52.4\degree, which corresponds to a dispersion of 1.53 $\lambda$/$D$.}
\end{figure}

A fixed ADC in the HCM has been defined so as to compensate for atmospheric dispersion at a moderate zenith angle of 32°. For any other zenith angle, this fixed configuration results with remaining atmospheric dispersion, with maxium effects at $5\degree \ $and $50\degree$. There will therefore be atmospheric dispersion residuals (from $-1.5\ \lambda/D$ at ZD$=5\degree \ $ to $1.3\ \lambda/D$ at ZD$=50\degree \ $ as shown in figure \ref{Dispersion}) that have to be taken into account in the measurements of the wavefront aberrations.      

\begin{figure}[H]
	\begin{center}
    	\includegraphics[width=\textwidth]{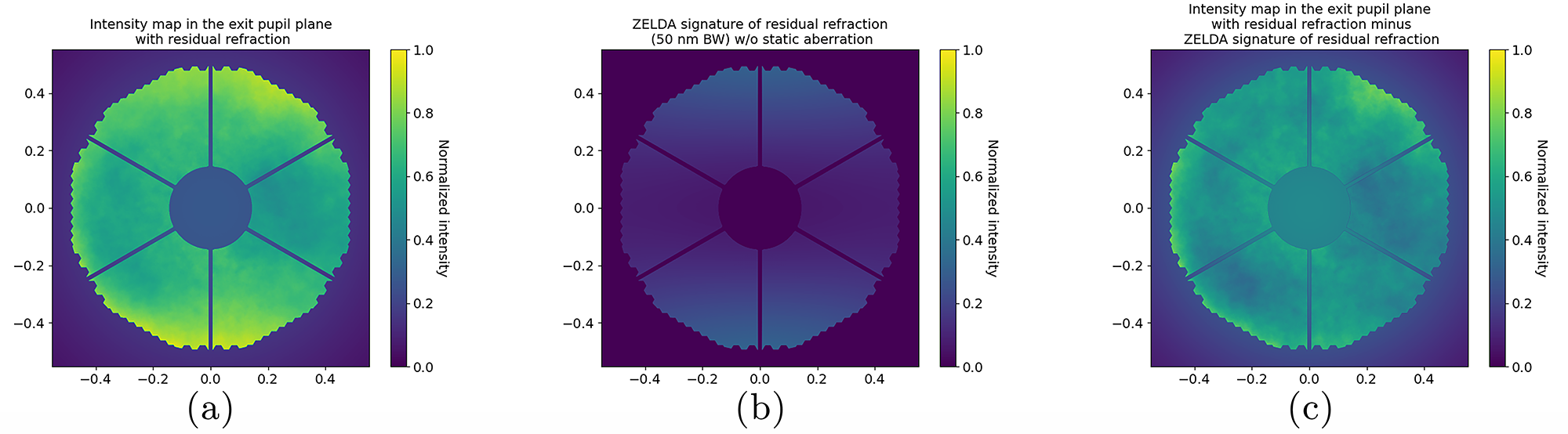}
	\end{center}
	\caption[Description liste figures]{\label{Disper} (a) ZELDA signature for a zenith distance of 52.4$\degree \ $ with a bandwidth of 50 nm. (b) Computed spurious signal induced by the atmospheric refraction residuals. (c) Estimated quasi-static signal by subtracting the spurious signal (b) from the ZELDA signature (a).}
\end{figure}

\noindent
The impact of atmospheric dispersion residuals on the ZELDA measurements depends on the deviation from the optimal observation angle and the considered bandwidth : in our tests (both numerical and experimental), we consider the conservative case beyond the limits of the HCM with a zenith distance of 52.4\degree. Finally, the ZELDA wavefront sensor works at a wavelength of 1.175 $\mu$m with a bandwidth of 50 nm. The different wavelengths will be dispersed along an axis: this will result in a PSF that will spread in one direction. Since the phase mask is designed to operate at 1.175 $\mu$m, the PSF of each scattered wavelength will only partially pass through the mask: this will result in a spurious signal resulting from the superposition of several tilted signals with respect to the signal of the central wavelength as shown in figure \ref{Disper}.

\noindent
This spurious signal must be taken into account to find the quasi-static aberrations that the ZELDA sensor must measure as shown by the results summarized in table \ref{Table3}.

\begin{table}[H]
\centering
\begin{tabular}{|r|c|c|c|c|} 
\hline
Spatial frequencies region & 0-4 $\lambda$/$D$ & 4-40 $\lambda$/$D$ & 40-100 $\lambda$/$D$ & ``all'' (0-512 $\lambda$/$D$)  \\ 
\hline
%$\varphi_{\text{out}}$ [nm rms]    & 10.91        & 7.30         & 2.22           & 13.88                       \\
%\hline
$\varphi_{\widetilde{\text{out}}}$ [nm rms] & 10.80 & 7.33 & 2.23 & 13.85 \\
\hline
%$\varphi_{\text{out-}\widetilde{\text{out}}}$ [nm rms] & 0.27 & 0.15 & 0.05 & 0.33 \\
%\hline
$\Delta\varphi_{\widetilde{\text{out}} \text{-uncalibrated dispersion}}$ [nm rms]    & 10.46        & 4.65         & 1.47           & 11.27                       \\
\hline
$\Delta\varphi_{\widetilde{\text{out}} \text{-calibrated dispersion}}$ [nm rms]    & 0.03        & 0.08         & 0.12           & 0.76                       \\
\hline
\end{tabular}
\caption[Description liste tableaux]{\label{Table3} Aberrations according to spatial frequencies in the input phase map ($\varphi_{\widetilde{\text{out}}}$) obtained with the Pupil D (see figure \ref{Pupil_approx}) and difference between uncalibrated/calibrated reconstructed and input maps.}
\end{table}

\noindent
As we can see, the spurious signal induced by atmospheric refraction residuals completely distorts the results. Our simulations have shown that a closed loop does not converge if this dispersion effect is not calibrated. This was tested experimentally and is detailed in subsection \ref{sec:part_3_3}.

\subsection{Adaptive Optics Residuals Effects}
\label{sec:part_2_4}

\begin{figure}[H]
	\begin{center}
    	\includegraphics[width=\textwidth]{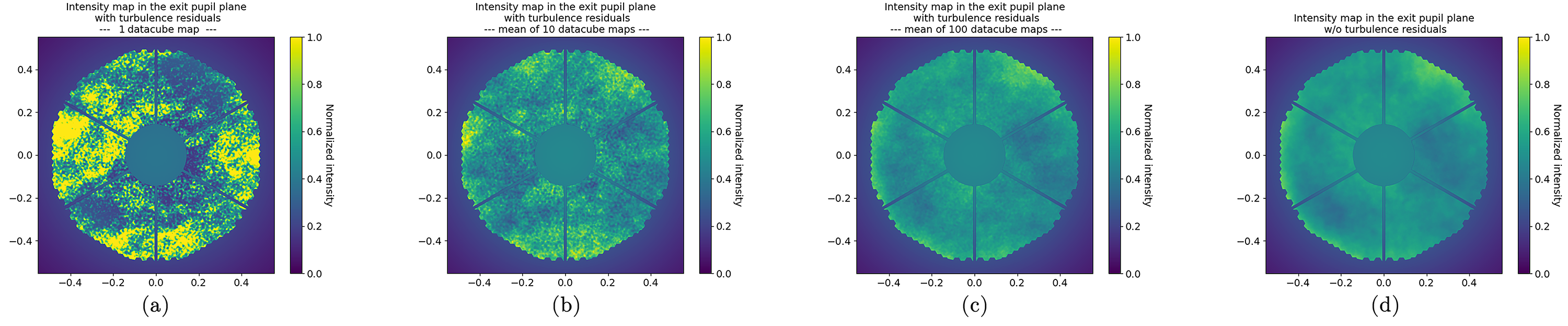}
	\end{center}
	\caption[Description liste figures]{\label{AO_res} ZELDA signature of turbulence residuals on top of static aberrations considering (a) only one map of the datacube, (b) the mean of 10 maps and (c) the mean of the all 100 maps of the datacube, all compared to the static aberrations only ZELDA signature (d).}
\end{figure}

The atmospheric turbulence will be corrected by the SCAO subsystem. However, this AO system is not as efficient as the extreme AO system of VLT-SPHERE, and there will be more turbulence residuals in the science path that will interfere with the ZELDA measurements. %Idem que avant : il n'existe pas de système parfait. On aura plus de problème dus à la turbulence que dans SPHERE, oui, mais SPHERE en avait quand même.
Typical turbulence residuals expected in HARMONI are of the order of 100 nm rms, which should theoretically outrange the ZELDA sensor which can only measure aberrations between -60 and 60 nm rms. Atmospheric turbulence varies rapidly over time, however, and, for a sufficiently large area, compensates for itself. If a wavefront is measured with the ZELDA sensor over a sufficiently long period of time, then the spurious signal induced by the turbulence residuals is averaged out and cancelled out, thus allowing the quasi-static aberration signal to be recovered. In order to compute this, we use a data cube containing 100 OPD maps computed from the same PSD, and we introduce the static aberrations map of figure \ref{Modele_Zelda}(a) added to the average map of $n$ maps of this data cube in the ZELDA simulator. Some results are shown in figure \ref{AO_res} and summarized in table \ref{Table4}. 

\begin{table}[H]
\centering
\begin{tabular}{|r|c|c|c|c|} 
\hline
Spatial frequencies region & 0-4 $\lambda$/$D$ & 4-40 $\lambda$/$D$ & 40-100 $\lambda$/$D$ & ``all'' (0-512 $\lambda$/$D$)  \\ 
\hline
$\varphi_{\widetilde{\text{out}}}$ [nm rms] & 10.80 & 7.33 & 2.23 & 13.85 \\
\hline
$\Delta\varphi_{\widetilde{\text{out}} \text{-}\overline{1}}$ [nm rms]    & 33.97        & 31.42         & 23.79           & 52.51                       \\
\hline
$\Delta\varphi_{\widetilde{\text{out}} \text{-}\overline{10}}$ [nm rms]    & 10.09        & 12.27         & 9.50           & 18.71                       \\
\hline
%$\Delta\varphi_{\widetilde{\text{out}} \text{-}\overline{25}}$ [nm rms]    & 8.30        & 8.14         & 6.09           & 13.25                       \\
%\hline
%$\Delta\varphi_{\widetilde{\text{out}} \text{-}\overline{50}}$ [nm rms]    & 5.19        & 5.46         & 4.22           & 8.73                       \\
%\hline
$\Delta\varphi_{\widetilde{\text{out}} \text{-}\overline{100}}$ [nm rms]    & 3.75        & 3.98         & 3.01           & 6.32                       \\
\hline
\end{tabular}
\caption[Description liste tableaux]{\label{Table4} Aberrations according to spatial frequencies in the input phase map ($\varphi_{\widetilde{\text{out}}}$) obtained with the Pupil D (see figure \ref{Pupil_approx}) and difference between reconstructed and input maps for different numbers of averaged maps ($\equiv$ different integration times).}
\end{table}

\noindent
For these simulations, we used 100 independent phase maps estimated to be similar to what is expected from HARMONI residuals and we considered the average of a certain number of these maps. This is an illustration of the expected effects but it is not a faithful representation because, in practice, the phase variations are not independent and talking about the number of phase maps is meaningless: it is the integration time that must be considered to average these effects of AO residuals. 

\noindent
According to table \ref{Table4}: the longer we average, the more accurate the measurement of static aberrations is. However, in practice, these aberrations are not static but quasi-static: the averaging time of the turbulence residuals is therefore limited by the lifetime of the quasi-static aberrations ($\sim$10-15s in HARMONI for the NCPA to change by a few nanometers \cite{carlotti:spie}).

%\subsection{Targeted Magnitudes}
%\label{sec:part_2_5}

%%% PART 3 %%%
\section{Experimental Investigation}
\label{sec:part_3}

As mentioned in the previous part, the ZELDA signal in the specific case of HARMONI is affected by :
\begin{itemize}
    \item The quasi-statics aberrations, which is the signal we want to measure, at $\lambda=$1.15-1.2 $\mu$m at a 0.06-0.1 Hz frequency, within a range of $\pm$60 nm rms and with an accuracy inferior or equal to 5 nm rms.
    \item The residual refraction which induces a spurious signal.
    \item Turbulence residuals which outrange the ZELDA sensor.
\end{itemize}

\noindent
In order to test the ZELDA in the specific case of HARMONI, we must therefore replicate the expected typical observation conditions in the most realistic way.

\subsection{Test Bench Description}
\label{sec:part_3_1}

\begin{figure}[H]
	\begin{center}
    	\includegraphics[width=\textwidth]{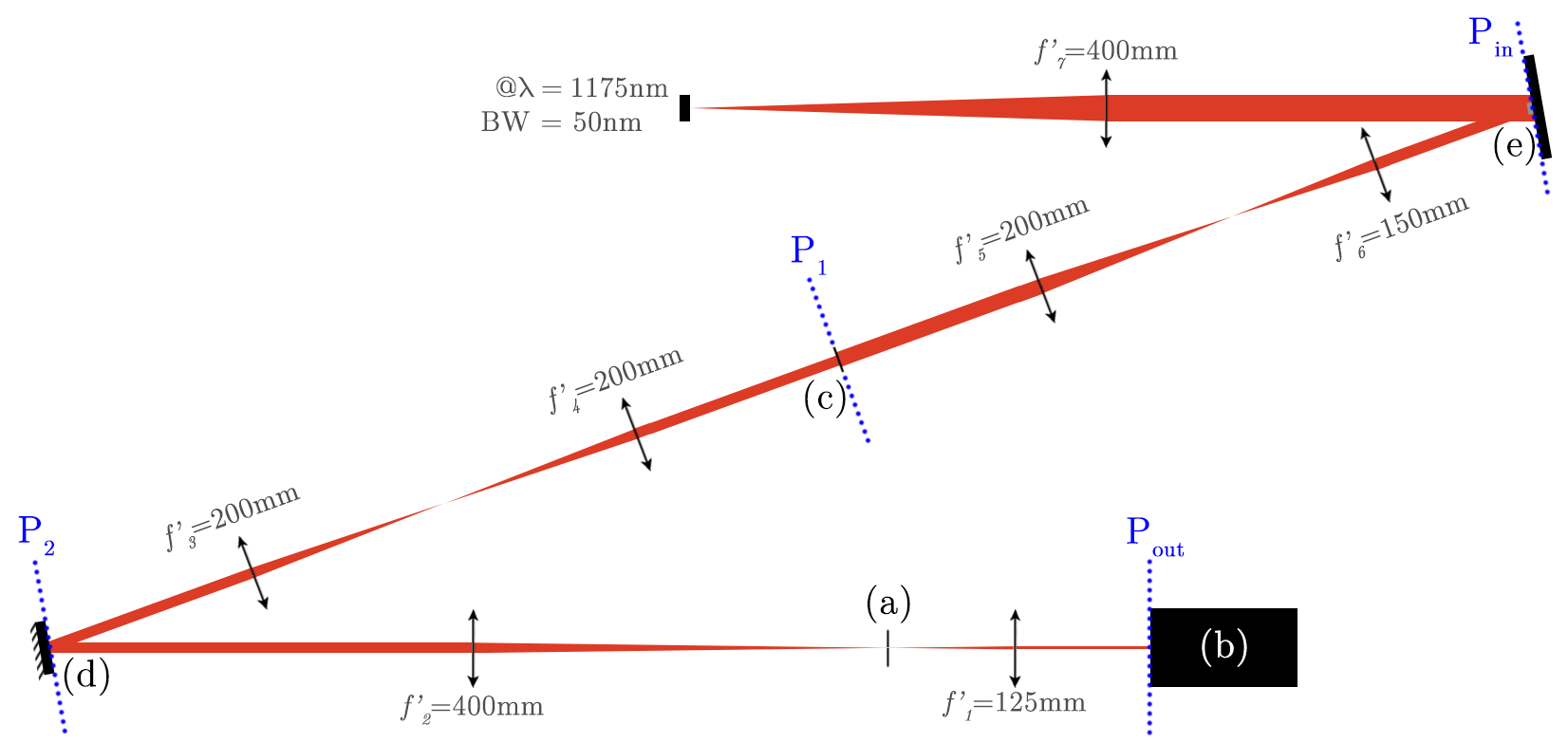}
	\end{center}
	\caption[Description liste figures]{\label{Bench} The experimental testbench to replicate HARMONI expected typical conditions. The ZELDA mask (a) is in a focal plane on a 3-axis motorized stages. The detector (b) is in the exit pupil plane P$_{\text{out}}$ where phase variations in the entrance pupil plane P$_{\text{in}}$ are converted into intensity variations. A $\sim$1/3900 scale replica of the ELT pupil (c) is in the pupil plane P$_{\text{1}}$ to have an aperture similar to the ELT one in the approximation of figure \ref{Pupil_approx} Pupil $D$. A prism (d) that introduce a dispersion of 1.53 $\lambda$/$D$ ($\sim$ zenith distance of 52.4$\degree \ $according to figure \ref{Dispersion}) is in the pupil plane P$_{\text{2}}$. A SLM (e) is in the entrance pupil plane P$_{\text{in}}$ and is used to mimic both typical turbulence residuals and static aberrations.}
\end{figure}

The experimental bench is described in figure \ref{Bench}. It is composed of the ZELDA phase mask from SILIOS, a prism, an ELT aperture replica, a Spatial Light Modulator (SLM) from Meadowlark, a C-RED 2 detector from First Light Imaging and a broadband source with a 50 nm bandwidth filter centered at 1.175 $\mu$m.

\begin{figure}[H]
	\begin{center}
    	\includegraphics[width=0.75\textwidth]{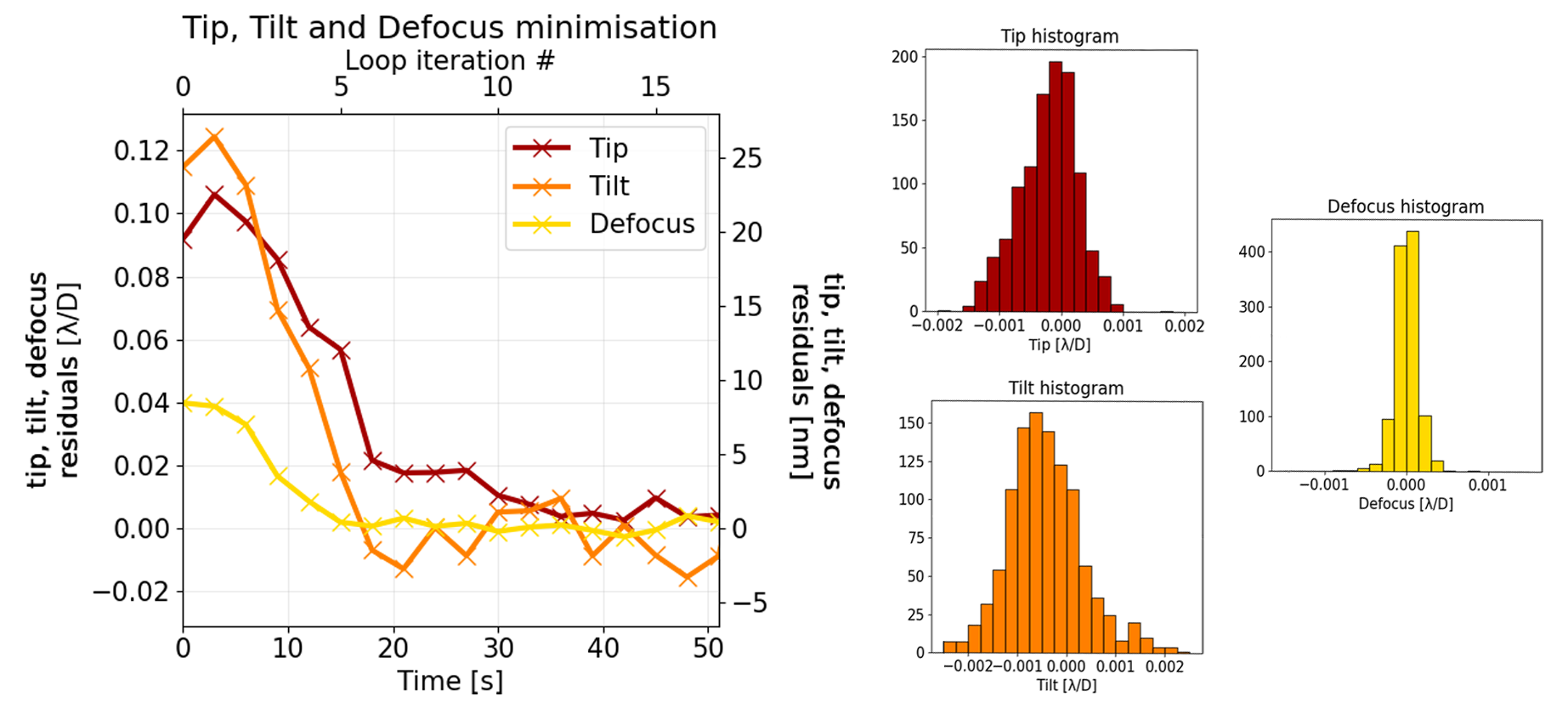}
	\end{center}
	\caption[Description liste figures]{\label{tip_tilt_def} Tip, tilt and defocus minimization by closing a loop on the 3-axis motorized PI stages.}
\end{figure}

\noindent
The phase mask is located in a focal plane on a set of 3-axis motorized PI stages controlled with Python. A closed loop can be established in order to center the mask and minimise tip, tilt and defocus to a subnanometric level as shown in figure \ref{tip_tilt_def}. This step of centring the ZELDA mask in the optical beam will be carried out with these same 3-axis motorized PI stages in the HCM\cite{jocou:spie}. The mask is designed to work at 1.175 $\mu$m with a f-number of f/40. The substrate is fused silica with a refractive index $n=1.4483$, and the mask has a diameter of $49.8 \pm 0.5$ $\mu$m (corresponding to $1.06\cdot\frac{\lambda f'}{D}$ at $\lambda=$ 1.175 $\mu$m, for a focal of 400 mm with a pupil diameter of 10 mm) and a depth of $655 \pm 5$ nm (checked by profilometry): it allows the phase mask to introduce a phase shift of $\frac{\pi}{2}$ and to convert the phase variation in the entrance pupil plane into intensity variations in the exit pupil plane where the detector is placed. The ZELDA sensor in an ideal case (without atmospheric refraction nor turbulence residuals) has been calibrated with a commercial wavefront sensor (a SID4 SWIR-HR from PHASICS) which operates in a 0.9 to 1.7 $\mu$m wavelength range, with $160 \times 128$ pixels, with a resolution of less than 2 nm rms and an accuracy of less than 5 nm rms. The difference between the measurements of the two wavefront sensors was around 3 nm rms for each tested OPD map, these 3 nm rms was due to the non-perfect surface of a folding mirror that was used to switch between the two sensors.

\noindent
The detector is a $640 \times 512$ InGaAs detector with 15 $\mu$m pixel pitch, is able to run at 600 FPS. It is driven with Python and is water-cooled to $-40\degree \ $with an external chiller.

\begin{figure}[H]
	\begin{center}
    	\includegraphics[scale=0.35]{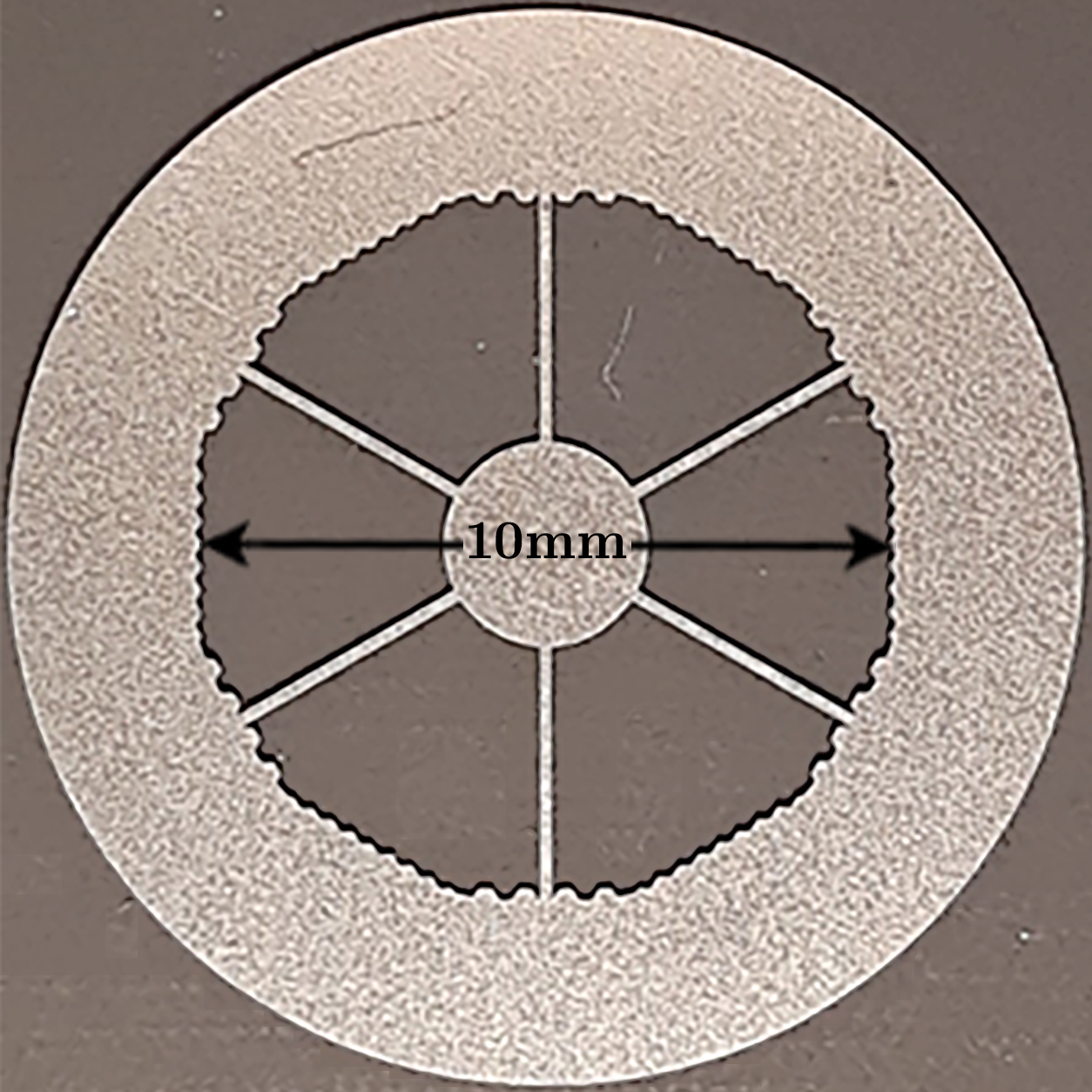}
	\end{center}
	\caption[Description liste figures]{\label{pup_banc} Laser-cutted amplitude mask to reproduce the ELT apperture shape approximation of figure \ref{Pupil_approx} pupil $D$.}
\end{figure}

\noindent
The telescope aperture shape of the ELT is replicated with a $\sim$1/3900 scale replica of the pupil $D$ approximation shown in figure \ref{Pupil_approx}. This physical pupil is shown in figure \ref{pup_banc}: the spiders are wider than the real ones due to the manufacturing constraints but the impact on the $b$ term and, consequently, on the phase estimation is negligible (OPD difference $\leq 0.2$ nm rms in the 0-512 $\lambda$/$D$ spatial frequencies range).

\noindent
The prism is used to mimic the atmospheric refraction residuals. It introduces a dispersion of 1.53 $\lambda$/$D$ ($\sim$ zenith distance of 52.4$\degree \ $according to figure \ref{Dispersion}) and is placed in a pupil plane. This prism allows us to test the  capability of the ZELDA sensor to measure the quasi-static aberrations in the pessimistic case of an observation beyond the limit of the HCM fixed at a zenith distance of 50\degree.

\noindent
The SLM is used to mimic both typical turbulence residuals and static aberration expected in the specific case of HARMONI. It is a $1920 \times 1152$ liquid crystals matrix which can introduce locally a phase shift up to 2$\pi$ according to the applied voltage. Its behavior is non-linear, and a Look-Up-Table (LUT) had to be experimentally constructed to drive the SLM correctly.

\noindent
The 3-axis motorized stage that supports the ZELDA mask, the detector and the SLM are driven with Python: it allowed the creation of an ``all-in-one'' user interface to fully control the experimental bench and allow communication between the detector and the SLM. We can thus correct in a closed loop the wavefront measured by the ZELDA sensor by sending a command to the SLM after having converted the intensity map received by the detector into an OPD map and then into an SLM command map via the SLM LUT. In addition to these conversions, it is also necessary to rescale the OPD map obtained with the detector so that the projection of the pupil on the SLM corresponds to the projection of the pupil on the detector: it is therefore necessary to measure upstream the decentering and the angle between the two arrays of pixels, as well as the x and y stretches (which are different due to the anamorphosis of the pupil since it is located in a pupil plane having an angle of about 8 degrees with the pupil planes of the SLM and of the detector).

\subsection{Close Loop In The Ideal Case -- Neither Atmospheric Refraction Nor AO Residuals}
\label{sec:part_3_2}

\begin{figure}[H]
	\begin{center}
    	\includegraphics[width=\textwidth]{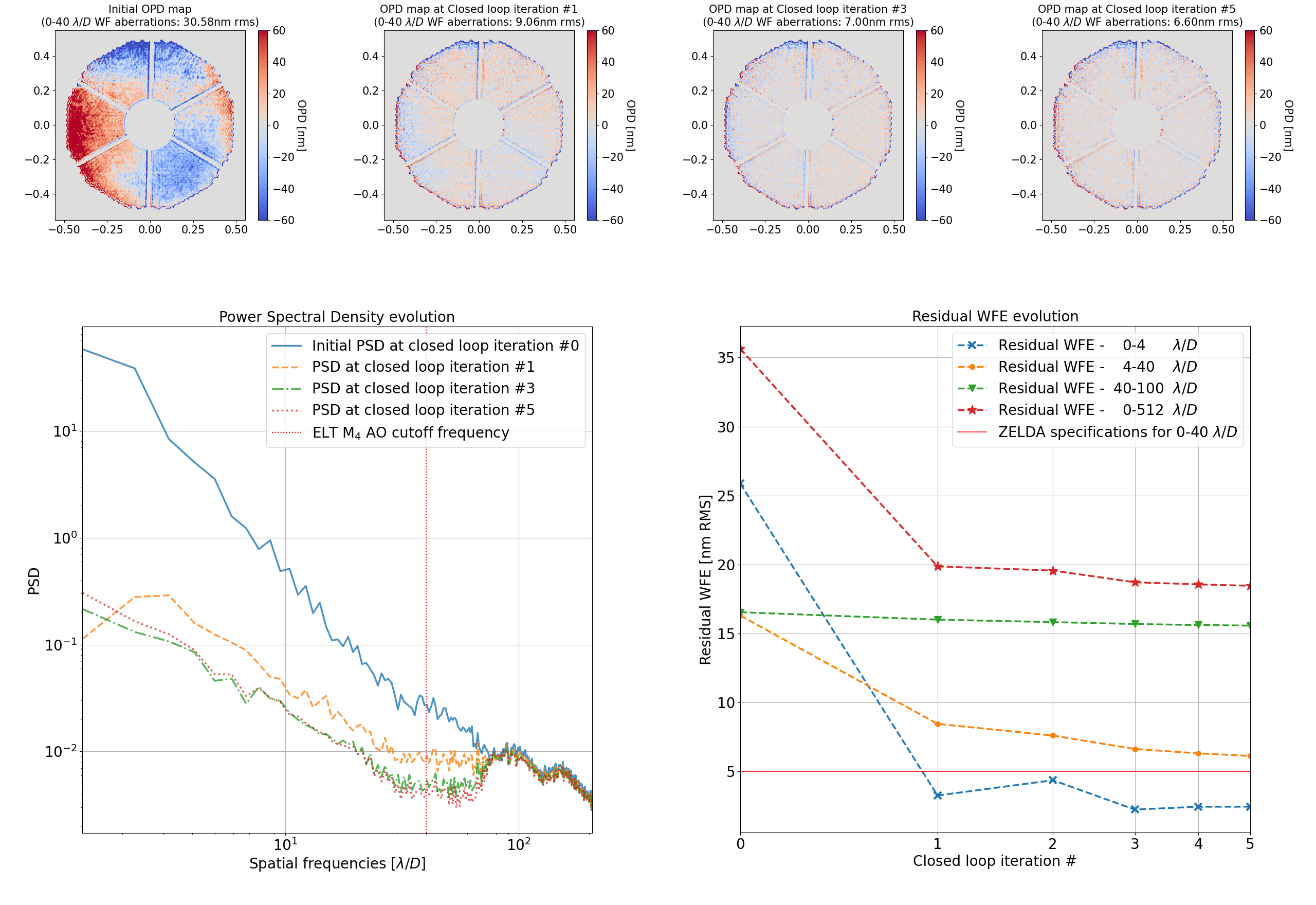}
	\end{center}
	\caption[Description liste figures]{\label{CL1} Evolution of the ZELDA signature, wavefront error residuals and PSD when closing a loop in the ideal case (neither dispersion nor AO residuals).}
\end{figure}

\noindent
The first closed loop correction was performed in the ideal case where there were neither atmospheric refraction nor AO residuals. Only the static aberrations of the experimental bench were measured by the ZELDA sensor. The results obtained are given in figure \ref{CL1}. 

\noindent
The correction on the bench is limited to 40 $\lambda$/$D$ as will be the correction on the ELT: for this, we filter the high spatial frequencies in Fourier space by multiplying the Fourier transform of the OPD map by a Hanning window before applying an inverse Fourier transform. The convergence of the correction loop is fast, so the SLM and the detector seem to communicate as expected and have been correctly calibrated. Another correction in the ideal case and up to 50 iterations was performed: the total residual WFE (0-512 $\lambda$/$D$) at the $50^{th}$ iteration was 10.08 nm rms, and 5.59 nm rms for the residual WFE between 0 and 40 $\lambda$/$D$. In this case, the loop is stable but the redundant WFE between 0 and 40 $\lambda$/$D$ does not reach the maximum 5 nm rms specified for HARMONI. One of the reasons that could explain this is a defect of scaling and pupil centering between the detector space and the SLM space: as can be seen on the OPD maps of figure \ref{CL1}, the pupil edges do not seem to be corrected. New tests in the ideal case will be performed soon to check this.

\noindent
Finally, these tests have also highlighted an optical gain on the measurements of the ZELDA sensor. Indeed, the ZELDA seems to underestimate the measured phase by a constant factor between 1.4 and 1.5. This factor was not detected during the calibration of the ZELDA with the commercial wavefront sensor PHASICS. Two reasons may explain this factor: 
\begin{itemize}
    \item An enclosure was added around the bench after the calibration and before these closed loop tests. The phase mask and the detector are very close to the edges of the enclosure. Light passes through circular openings, where microturbulences may interfere with the wavefront measurements (see equation \ref{terme_b}).
    \item The mechanical support of the SLM is particularly unstable, with a consistent temporal drift of the tilt ($\sim \pm 4$ nm/min). This mechanical instability may be the result of a local thermal expansion, despite the fact that the SLM is water-cooled with an external chiller.  
\end{itemize}

\noindent
A test to verify if it is indeed these potential microturbulences that cause this optical gain will soon be carried out: we will observe the evolution of the PSF with very short exposure times and then with long exposure times in order to check if it moves on the detector. If it is the case, the long exposures will show a wider PSF compared to the PSF obtained at short exposure times, and this will highlight the presence of these turbulences.

\subsection{Close Loop -- Atmospheric Refraction Residuals On Top Of Static Aberrations}
\label{sec:part_3_3}

\begin{figure}[H]
	\begin{center}
    	\includegraphics[width=\textwidth]{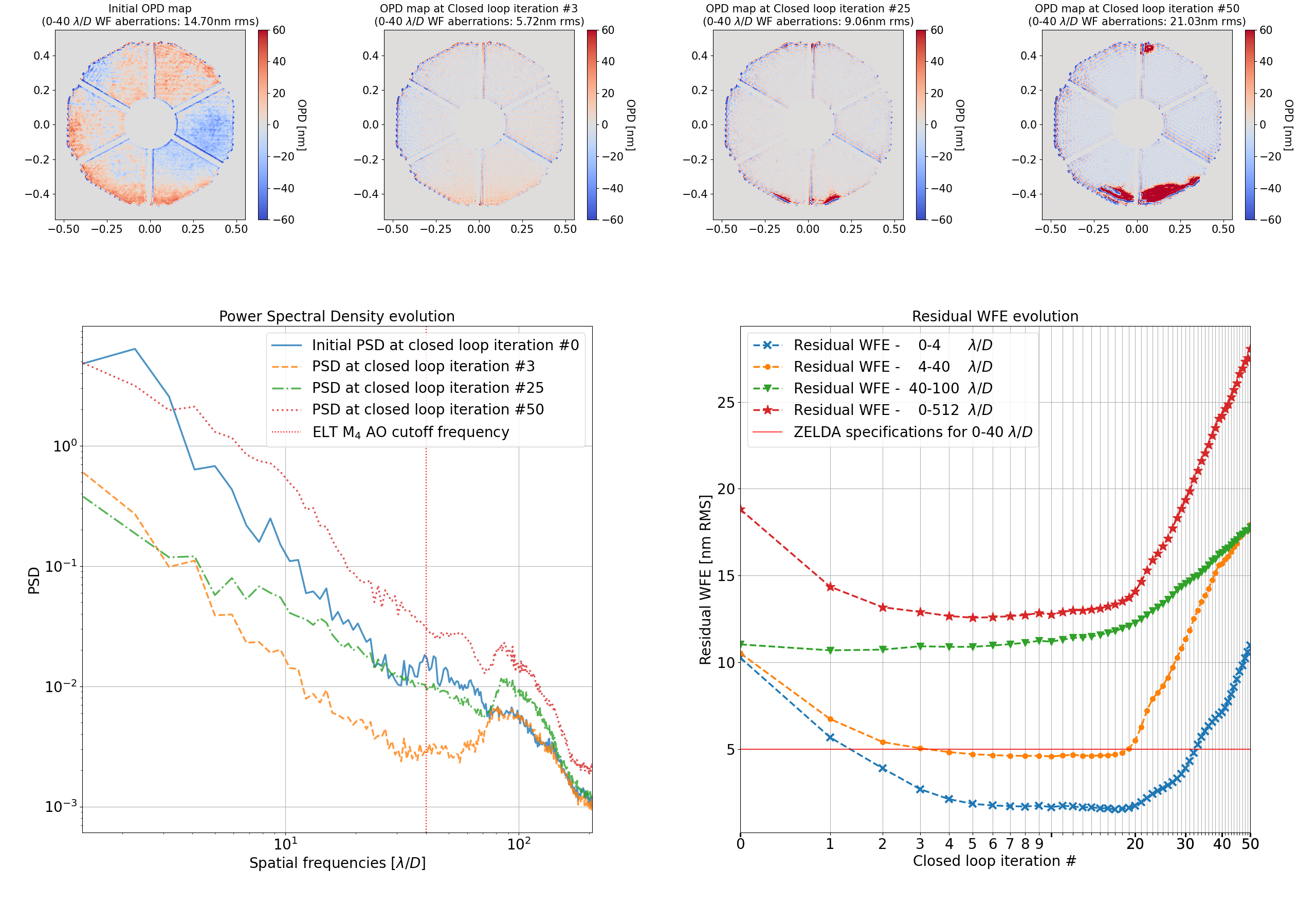}
	\end{center}
	\caption[Description liste figures]{\label{CL2} Evolution of the ZELDA signature, wavefront error residuals and PSD when closing a loop without calibrating the dispersion effect.}
\end{figure}

The closed loop test was then performed in a configuration where atmospheric refraction residuals were added to the static aberrations of the experimental bench. Consistent with the simulations presented in section \ref{sec:part_2_3}, the loop diverges if the dispersion effect is not calibrated, as shown in figure \ref{CL2}.

\noindent
This shows that the calibration procedure efficiently corrects for the spurious signal induced by the residual atmospheric dispersion.

\begin{figure}[H]
	\begin{center}
    	\includegraphics[width=\textwidth]{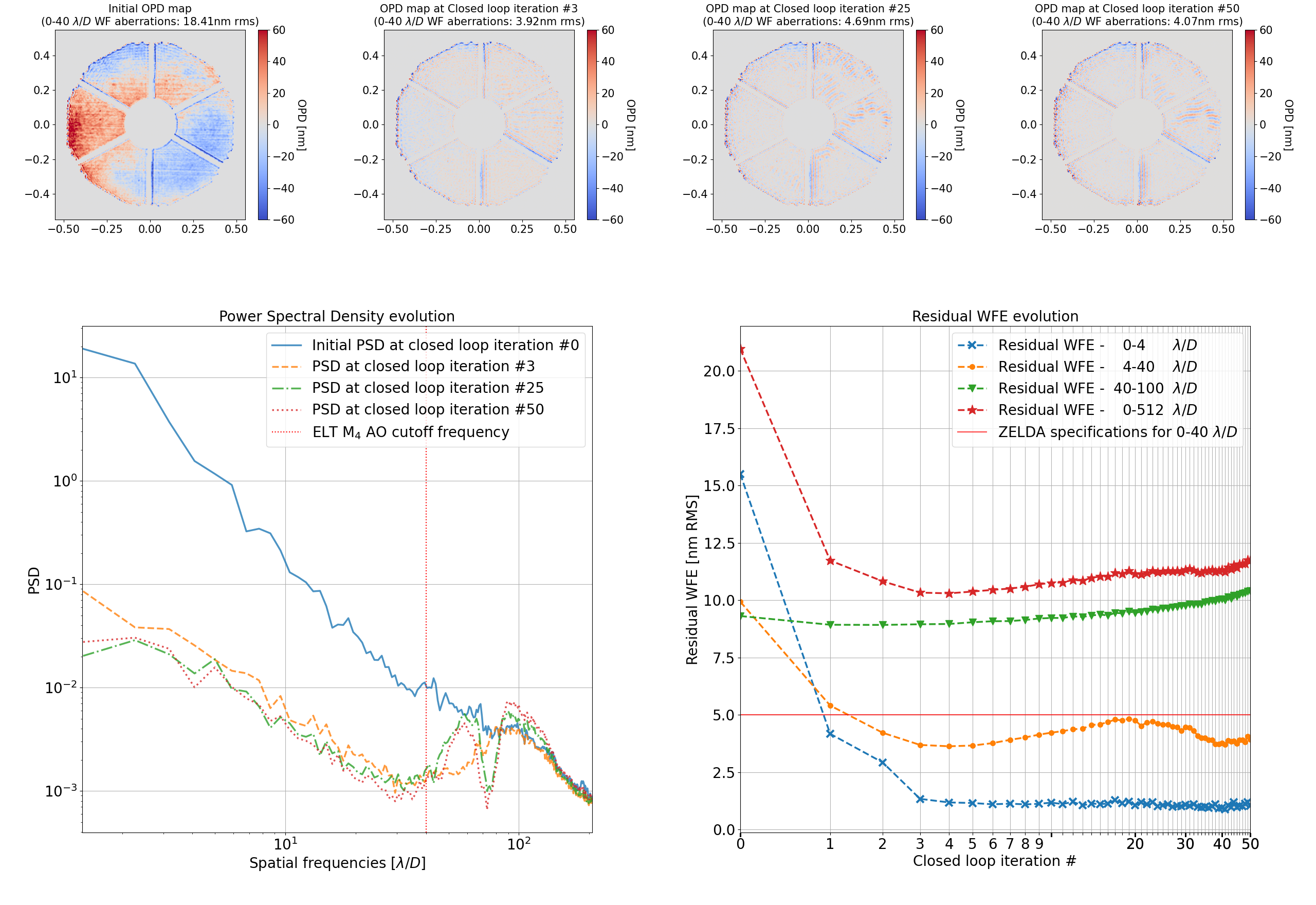}
	\end{center}
	\caption[Description liste figures]{\label{CL3}Evolution of the ZELDA signature, wavefront error residuals and PSD when closing a loop by calibrating the dispersion effect.}
\end{figure}

\noindent
If the effect is calibrated, the loop converges again, as shown in figure \ref{CL3}.

\noindent
As in the previous case, the correction on the bench is limited to 40 $\lambda$/$D$. Finally, the optical gain on the ZELDA sensor increases from 1.4-1.5 to 2 when the prism is added to the bench. The reason for this is still uncertain and will be investigated in the near future. 

\subsection{Close Loop -- AO Residuals On Top Of Static Aberrations}
\label{sec:part_3_4}

The closed loop correction with AO residuals introduced on the bench with the SLM is currently in progress. The first results, still very preliminary, seem to show a slower convergence of the loop than in the previous cases, and more advanced tests on a larger number of iterations will have to be performed to draw more solid conclusions.

%\begin{figure}
%	\begin{center}
%    	\includegraphics[width=\textwidth]{Images/Ao_res_results.png}
%	\end{center}
%	\caption[Description liste figures]{\label{CL4}...}
%\end{figure}

%\lipsum[1]

\section{CONCLUSIONS}
\label{sec:conclu}

The tests of the ZELDA sensor under the typical observing conditions expected with HARMONI are encouraging: the experimental results are consistent with the simulations, and the main obstacles to the use of a ZELDA wavefront sensor (presence of atmospheric dispersion residuals and Adaptive Optics residuals) are well controlled. Closed-loop tests with AO residuals are in progress. Tests combining the effects of chromatic dispersion and AO residuals, thus reproducing the HARMONI conditions as closely as possible, will soon be performed.

\noindent
The issue of optical gain between the calibration of the ZELDA sensor and the closed loop tests will also be investigated before the end of 2023.

\noindent
As a result of these tests and results, the ZELDA sensor appears to be ready to operate as expected under HARMONI conditions: atmospheric refraction and AO residuals do not prevent the measurement of NCPAs. %Achieving a contrast of $10^{-6}$ is therefore very likely.

%https://filesender.renater.fr/?s=download&token=b7cf9f66-aaf4-47a6-87a7-cc309bbff429

%\lipsum[1]

\appendix    %>>>> this command starts appendixes

%\section{MISCELLANEOUS FORMATTING DETAILS}
%\label{sec:misc}

%\subsection{Formatting Equations}

%\subsection{Formatting Theorems}

%\acknowledgments % equivalent to \section*{ACKNOWLEDGMENTS}       

% References
\bibliography{main} % bibliography data in report.bib

\begin{thebibliography}{1}

\bibitem{Mamadou2016}
{N\'{}Diaye, M.}, {Vigan, A.}, {Dohlen, K.}, {Sauvage, J.-F.}, {Caillat, A.},
  {Costille, A.}, {Girard, J. H. V.}, {Beuzit, J.-L.}, {Fusco, T.}, {Blanchard,
  P.}, {Le Merrer, J.}, {Le Mignant, D.}, {Madec, F.}, {Moreaux, G.},
  {Mouillet, D.}, {Puget, P.}, and {Zins, G.}, ``Calibration of quasi-static
  aberrations in exoplanet direct-imaging instruments with a zernike phase-mask
  sensor - ii. concept validation with zelda on vlt/sphere,'' {\em A\&A}~{\bf
  592},  A79 (2016).

\bibitem{carlotti:hal-02118132}
Carlotti, A., H{\'e}nault, F., Dohlen, K., Sauvage, J.-F., Rabou, P., Magnard,
  y., Vigan, A., Mouillet, D., Chauvin, G., Vola, P., Fusco, T., El~Hadi, K.,
  Thatte, N., Clarke, F., Tecza, M., Bryson, I., Schnetler, H., and
  V{\'e}rinaud, C., ``{System analysis and expected performance of a
  high-contrast module for HARMONI},'' in [{\em {SPIE Astronomical Telescopes +
  Instrumentation}}{\nolinebreak\hspace{0.1em}]},   {\bf 0702},  352, {SPIE},
  Austin, United States (June 2018).

\bibitem{Mamadou2013}
{N\'{}Diaye, M.}, {Dohlen, K.}, {Fusco, T.}, and {Paul, B.}, ``Calibration of
  quasi-static aberrations in exoplanet direct-imaging instruments with a
  zernike phase-mask sensor,'' {\em A\&A}~{\bf 555},  A94 (2013).

\bibitem{carlotti:spie}
Carlotti, A. and {et al.}, ``{HARMONI at ELT: System analysis and performance
  estimation of the high-contrast module},'' in [{\em {SPIE Astronomical
  Telescopes + Instrumentation}}{\nolinebreak\hspace{0.1em}]},   {\bf
  12185-211},  AS22--AS106--96, {SPIE}, Austin, United States (July 2022).

\bibitem{jocou:spie}
Jocou, L. and {et al.}, ``{HARMONI at ELT: development of the High Contrast
  module},'' in [{\em {SPIE Astronomical Telescopes +
  Instrumentation}}{\nolinebreak\hspace{0.1em}]},   {\bf 12185-171},
  AS22--AS106--8, {SPIE}, Austin, United States (July 2022).

\end{thebibliography}
\bibliographystyle{spiebib} % makes bibtex use spiebib.bst

\end{document}